\documentclass[twocolumn,showpacs,preprintnumbers]{revtex4}
\usepackage{epsfig}
\usepackage{graphics}
\usepackage{subfigure}

\begin{document}

\def\Huge{\huge}
\def\e{\begin{equation}}
\def\f{\end{equation}}
\def\*{^{\displaystyle*}}
\def\=#1{\overline{\overline #1}}
\def\d{\partial}
\def\s{\strut\displaystyle}
\def\-#1{\vec #1}
\def\o{\omega}
\def\O{\Omega}
\def\va{\varepsilon}
\def\D{\nabla}
\def\.{\cdot}
\def\x{\times}
\def\l#1{\label{eq:#1}}
\def\r#1{(\ref{eq:#1})}
\def\am{\left(\begin{array}{c}}
\def\amm{\left(\begin{array}{cc}}
\def\a{\end{array}\right)}
\newcommand{\ds}{\displaystyle}

\title{Metamaterial nanotips}

\author{C. Rockstuhl$^1$, C. R. Simovski$^2$, S. A. Tretyakov$^2$ and F. Lederer$^1$ }
\affiliation{$^1$ Institute of Condensed Matter Theory and Solid
State Optics, Friedrich-Schiller-Universit\"{a}t Jena, Germany,
\\
$^2$ Helsinki University of Technology, Finland}

\begin{abstract}
Nanostructured metamaterials, especially arrays of metallic nanoparticles which
sustain the excitation of localized plasmon polaritons, provide excellent
opportunities to mold the flow of light in the linear regime. We suggest a
metamaterial structure whose properties are determined not only by its inner
geometry but also by its entire shape. We call this structure a
\emph{metamaterial nanotip}. We evaluate the potential of this nanotip to
control the size and the location of the field enhancement. Two-dimensional
implementations of this metamaterial nanotip were comprehensively numerically
simulated and confirm the expected, physically distinct regimes of operation.
\end{abstract}

 \maketitle

Abundant literature is devoted to the enhancement of local light
fields using metallic nanoparticles. This effect is the basis of
surface enhanced Raman scattering (SERS) (see e.g. \cite{11,13,4}).
Nowadays it is also used for localizing and detecting different
nanoobjects, including molecules (see e.g. \cite{5,6,7}), and even
for the control of their fluorescence \cite{8,9}. Exciting results
(local field enhancement by a factor of $10^2$) were obtained for
nanoantennas realized as two closely spaced metallic nanoparticles
\cite{10}. In extremely narrow gaps ($1\dots 2$ nm) between two
almost touching metallic nanoparticles the local field can grow even
more significantly \cite{12}.  In arrays of touching nanospheres
with rather large diameters (when the plasmonic resonance is
multipolar) the field enhancement in the $1$ nm sized domain near
the contact point may attain $10^6$ \cite{15}. It explains
extraordinary experimental data for the Raman scattering enhancement
(up to $10^{14}$) obtained in Ref.~\onlinecite{16}.

However, in some applications a trade-off between a maximal local
field enhancement and a maximal area where the field is enhanced
appears. In fact, in SERS studies situations often arise where the
molecules to be detected float in a liquid or a gas \cite{review}.
It is then difficult to steer these molecules into a $1$ nm sized
gap of an optical nanoantenna. In such situations it would be more
favorable to create larger spot sizes (e.g. $50\dots 100$ nm) at the
expense of a reduced field enhancement. Metamaterials (MM) based on
metallic nanoparticles open the opportunity to realize and to tailor
such spots. As it was shown, e. g., in Ref.~\onlinecite{PRL}, arrays
of small (a few nm) metal nanoparticles can be used to induce a
strong dispersion of the effective permittivity with a Lorentzian
line shape. The resonance is centered at that wavelength where the
localized plasmon polariton is excited in the ensemble of metal
nanoparticles. Such MM permits to accomplish a cavity from closely
spaced metal nanoparticles having an effective permittivity not
achievable with natural occurring materials. Choosing a proper
design for the cavity and operating it in different spectral domains
permits to control the location, where light localizes, as well as
the field area and magnitude.

The purpose of this paper is to design a MM cavity that concentrates
the incident light in a hot spot whose size and shape can be
efficiently and reliably controlled by the design parameters of the
MM. Two options for the hot spot are considered, namely being
located inside or outside the cavity. Prior to further
considerations we expect that our proposed MM cavity is subject to
an optical analogue of the uncertainty principle. Namely, field
enhancement and spot size remain intrinsically coupled. For example,
it will be impossible to obtain a huge field enhancement in large
spots. However, the intriguing advantage of the proposed MM cavity
is that both quantities can be adjusted to the largest possible
extent suggesting many potential applications. Examples are elements
to couple light from propagating waves to optical nanocircuits or
nanofilters \cite{Engheta}, to obtain controlled \emph{photonic
nanojets} \cite{Taflove,Taflove1}, as extremely robust tips for
scanning near-field optical microscopes, or for biosensors.

We propose a MM cavity of submicrometer size implemented as a small
cone made of metal nanoparticles. We call this structure a \emph{MM
nanotip}. The size of the cone's base can be of the order of
$\lambda/2$ or larger. A solid immersion lens or a conventional
microcavity creating the conventional light nanojet can be used for
focusing the primary light beam on top of the MM nanotip. The MM
nanotip further squeezes the nanojet. The coupling of the MM cavity
to the incident wave beam is strong since the incident light
transmits through the flat cone base. The location of hot spots,
their size and shape strongly depend on the MM design parameters
(the distances between nanoparticles, their shape, and the size of
the cone). Another control parameter is the cone truncation;
transforming it to a frustum. Furthermore, the effective
permittivity of the cone is determined by the array of nanoparticles
and depends only weakly on the matrix. This provides unique features
of the structure. First, the nanotip does not need to have a conical
physical geometry. By creating a conically shaped array of
nanoparticles in a dielectric slab (if we ensure that the apex of
the cone coincides with an interface of the slab), effective field
localization occurs around a singular surface point at the bottom of
the slab. Second, the MM nanotip can be penetrable for molecules
(the matrix can be porous, alternatively one can make voids in the
matrix). Both properties are useful for applications indicated
above.

\begin{figure}
\begin{center}
\includegraphics[width=85mm]{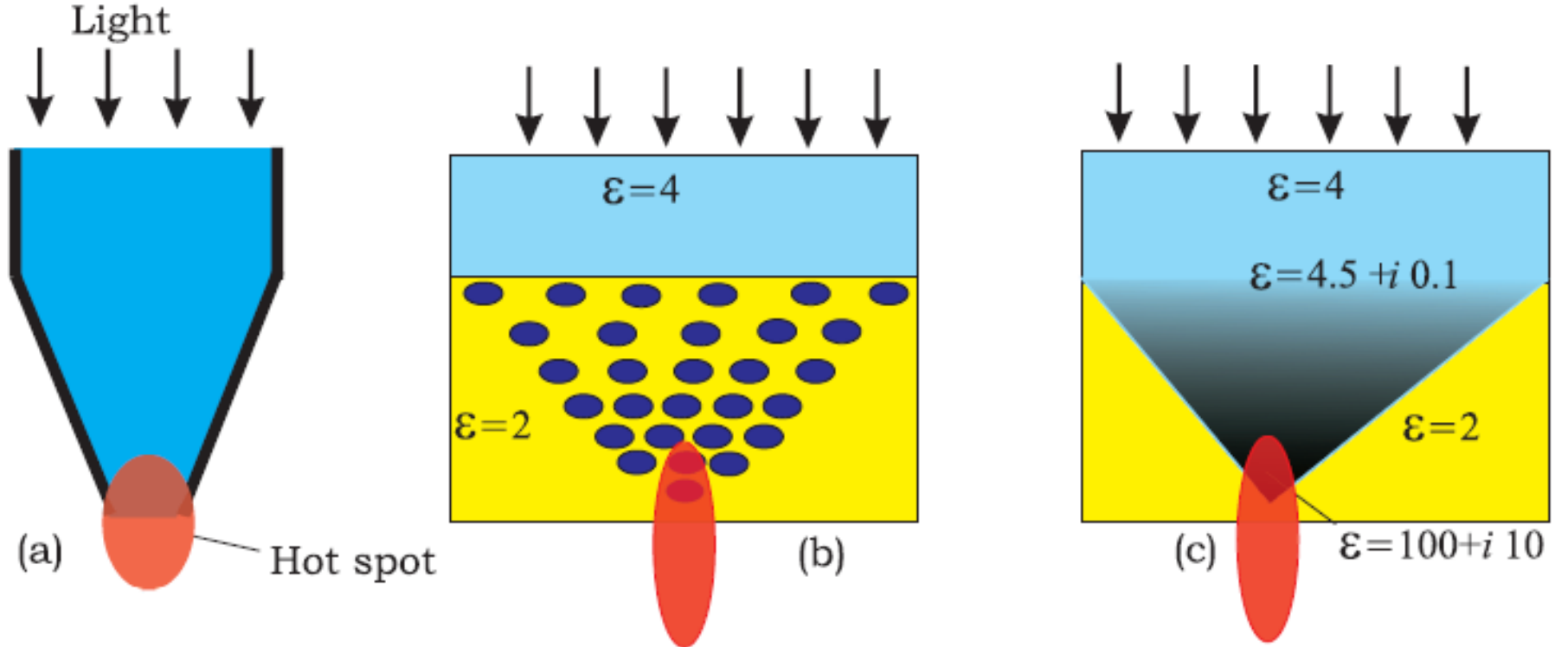}
\caption{(Color online) (a) A tapered SNOM tip with a metal coating.
(b) A MM nanotip with gradually increasing concentration of
nanoparticles. (c) The effective medium model of the MM nanotip
explaining the reflection suppression. The values of the
permittivity of the upper and lower dielectric matrices as well as
the values of the effective permittivity shown on top and bottom of
the MM nanotip are purely illustrative.} \label{fig1}
\end{center}
\end{figure}

High reflection from the cone base can be prevented by gradually
increasing the concentration of nanoparticles from the base to the
apex of the cone. The light will  experience an effective medium
with a permittivity increasing without jumps between the base and
the apex thus minimizing the reflection losses. In Fig.~\ref{fig1} a
sketch of the MM nanotip together with a conventional tapered
metal-coated SNOM tip with a subwavelength aperture is displayed.
The figure illustrates the idea of the engineered high permittivity
cone, our design strategy, and the regime of the nanojet. We
emphasize that in this way one can achieve a wave impedance of the
effectively tapered transmission line approximately uniform along
the entire tip length. As a consequence this MM nanotip may be
potentially employed to efficiently couple light into nanofilters,
waveguides of metallic nanoparticles \cite{Atwater} and
nanoantennas.

At this first stage we simulate a 2D analogue of the suggested
structure. Instead of nanoparticles we assume here nanocylinders.
For the sake of saving computational time we consider a regular
array i.e. no pitch (ratio period/diameter) variation along the
nanotip yet. With this approach we cannot reduce spurious
reflections at the base, though these reflections do not affect the
main conclusions to be drawn and leaves this optimization for a
future work. Therefore, the structure consists of periodically
arranged metallic nanowires forming the tip. As metal material we
have chosen silver. Material parameters were taken from literature
\cite{Christy}. The dielectric matrix is assumed to be glass
($n=1.5$). The diameter of the silver cylinders was $D~=~4~$nm. The
shape of the tip was chosen to be an equilateral triangle.
 The structure was illuminated by a TM polarized (electric field
 in the plane of incidence ) plane wave with
unit amplitude. The main parameters, modified in this work to identify the
operational regimes of interest, are the wavelength and the period.

The effective permittivity of the MM made of closely spaced
cylinders was obtained rigorously from the dispersion
relation\cite{Dispersion_1}. The MM was reasonably assumed to have
no spatial dispersion in the effective permeability. The dispersion
relation was calculated with a appropriate plane wave expansion
technique. The field distribution around the ultimate device was
computed by using an extended Mie theory to treat the case of light
scattering at an ensemble of non-penetrating cylinders.

To start with, Fig.~\ref{fig2} shows the effective permittivity of
the MM made of nanocylinders as a function of wavelength and for
various periods. It is evident, that strong dispersion with a
Lorentzian line shape occurs near the wavelength where the localized
plasmon polariton is excited in the nanocylinder. The smaller the
period the larger is the resonance wavelength. The red-shift arises
from the mutual coupling of the resonant fields in adjacent
cylinders. This coupling and the larger filling fraction causes also
the larger resonance strength for smaller periods. Two spectral
domains of interest can be distinguished in general. For wavelengths
below resonance the MM behaves metallic
($\Re(\epsilon_\mathrm{Eff})~<~0$) whereas it has dielectric
properties ($\Re(\epsilon_\mathrm{Eff})~>~1$)beyond the resonance.
We note that the effective permittivity exceeds by far that of any
naturally occurring material at optical frequencies. Both regimes
can be used to localize light. In the 'metallic' domain, surface
plasmon polaritons can be excited in the tip. They will cause a
strong field concentration at the tip apex in small volumes. In the
'dielectric' domain whispering gallery modes are excited in the tip
leading to a diffraction-limited field concentration inside the tip.
The diffraction limit is given by the ratio of the wavelength to the
huge effective refractive index. If the operational wavelength is
much larger than the resonance wavelength the field focus may be
pushed off the tip. This leads to a strong field enhancement in
close vicinity but outside the tip. This operational regime, where
the formation of photonic nanojets can be observed, adds as a third
one to those discussed above.

\begin{figure}
\begin{center}
\includegraphics[width=85mm]{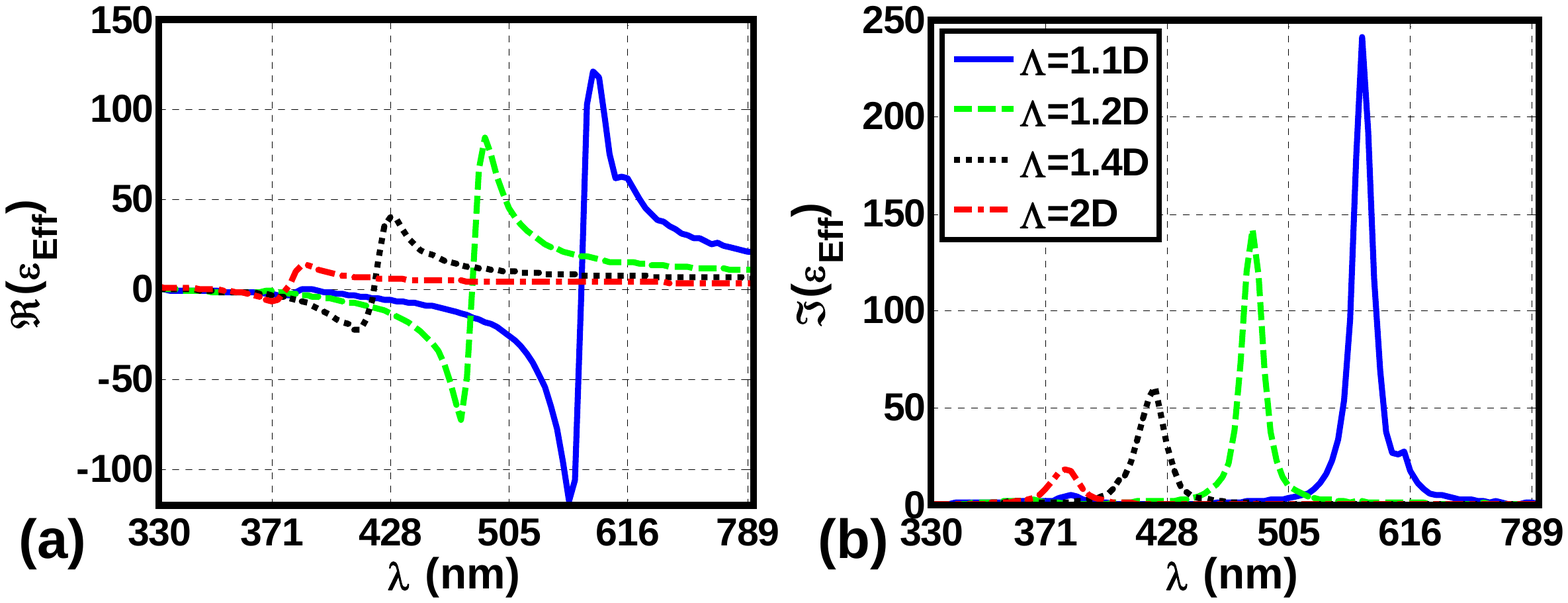}
\caption{(Color online) Real (a) and imaginary (b) part of the
effective permittivity as a function of the wavelength for various
periods and a MM made of nanocylinders (diameter $D = 4$ nm)
arranged on a square grid.} \label{fig2}
\end{center}
\end{figure}

To identify spectral and parameter domains of interest, the
electromagnetic field was computed close to the apex of the tip. The
parameters we systematically investigated were the period $\Lambda$
and the size of the tip. Selected results for various periods are shown in
Fig.~\ref{fig3}. Strong enhancement and field localization can be
seen. This data have been used to extract the parameters space of interest where any of
the operational regimes defined are fully evolved.

\begin{figure}
\begin{center}
\includegraphics[width=85mm]{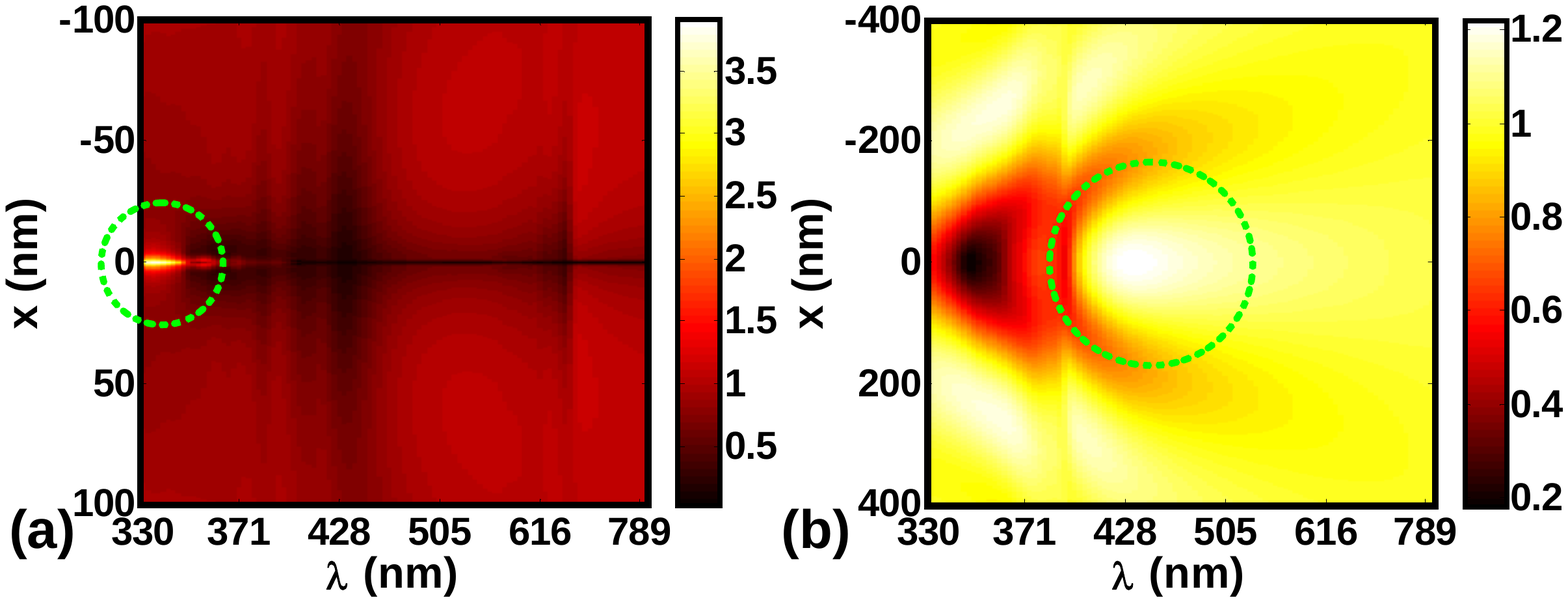}
\caption{(Color online) Electric field amplitude as a
function of the wavelength and the transverse coordinate $x$ at $z=1$ nm (a) and $z=70$ nm (b)
behind the apex ($x~=~0$~nm) of the tip made of nanocylinders. In both cases the period is $\Lambda~=~2D$
and the tip is made of 20 nanocylinder layers. The interesting
parameter domains with strong field enhancements are indicated by
circles.} \label{fig3}
\end{center}
\end{figure}

Finally we have analyzed the entire field distribution for a
selected number of relevant configurations. Results are shown in
Fig.~\ref{fig4}.

For the sake of comparison Fig.~\ref{fig4} (a) shows the electric
field amplitude around a single nanocylinder. The chosen wavelength
corresponds to the plasmon polariton resonance.  Near the surface a
field enhancement $\chi=|E^{\rm loc}/E_0|$ of up to $\chi=15$ can be
observed. However, 1 nm off the cylinder surface it decreases to
$\chi=6$ and the averaged enhancement over the area $20\time 20$ nm
off the cylinder does not exceed $\chi=1.1$.

Figure~\ref{fig4} (b) shows exemplarily a field distribution in the
operational regime where the effective properties of the tip
material are metallic and surface plasmon polaritons are excited in
the tip. The field enhancement is strong at any apex and attains a
maximum of $\chi=7$. Reminiscent to the plasmonic properties of the
entire tip is the rather constant field amplitude inside the tip,
being comparable to the field amplitude of the single nanocylinder
as shown in Fig.~\ref{fig4} (a).

Figure~\ref{fig4} (c) shows a field distribution in the operational
regime where the effective properties of the tip material are
essentially dielectric (only a minor imaginary part of the effective
permittivity) and whispering gallery modes are excited. The field
enhancement $\chi$ is strong at three hot spots inside the tip and
can be further increased for $\Lambda\rightarrow D$.

Figure \ref{fig4} (d) shows exemplarily a field distribution in the
third regime where the effective properties of the tip material are
essentially dielectric (negligible imaginary part of the effective
permittivity) but the field is dragged out of the tip and a photonic
nanojet appears. The amplitude enhancement inside the nanojet
($\chi\approx 1.7$) is a sign for an excellent optical coupling from
the incident wave to the photonic nanojet. The effective width of
the nanoject is approximately $120$ nm ($0.42\lambda_h$, where
$\lambda_h=\lambda/n$ is the wavelength in the host medium). We note
that due to the wedge shape, the structure resembles a
two-dimensional axicon. In a first approximation the evolving beam
has two major spatial frequencies. The wedge shape structure
therefore significantly suppresses diffraction, being much in favor
for the photonic nanojets.

\begin{figure}
\begin{center}
\includegraphics[width=85mm]{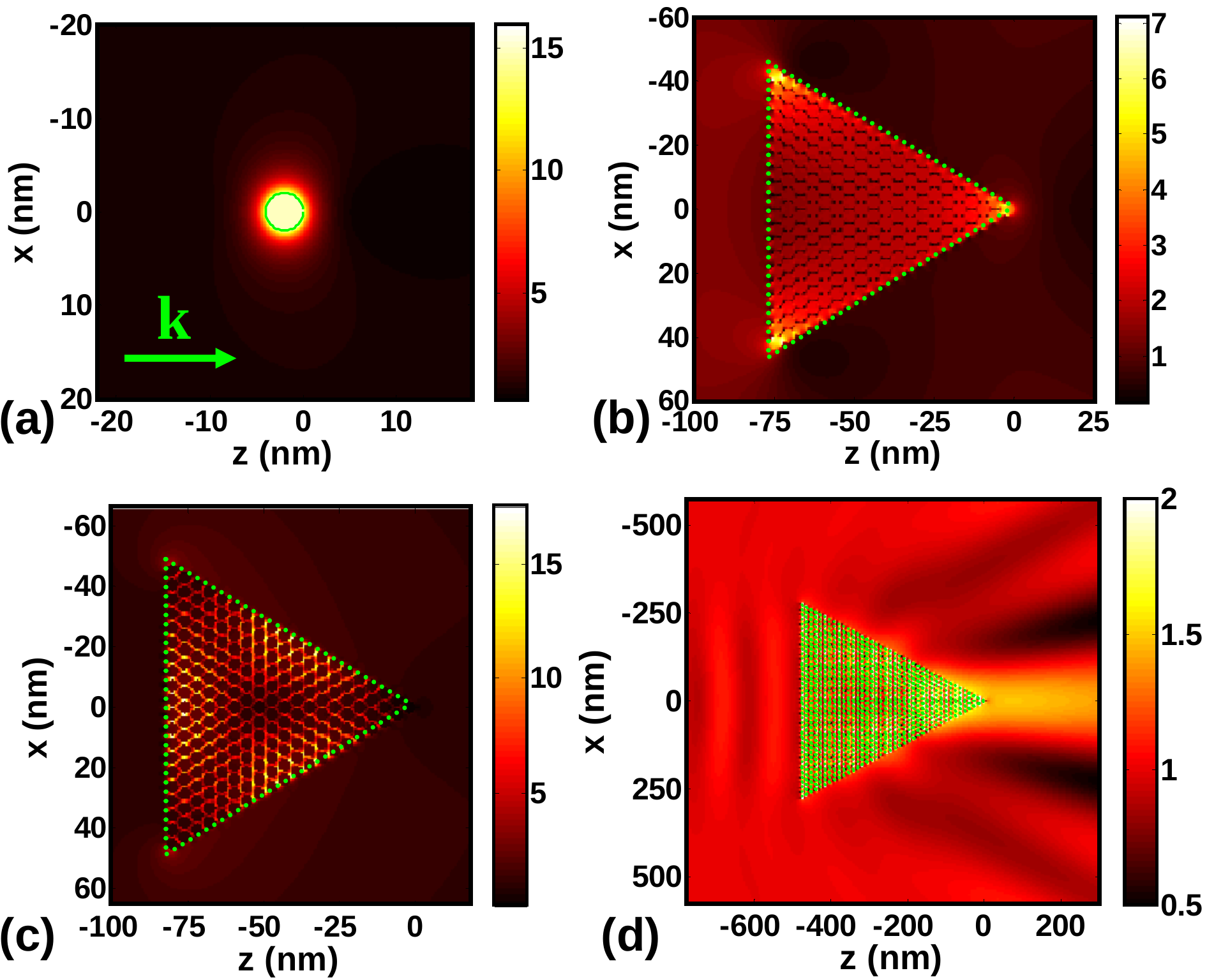}
\caption{(Color online) (a) Electric field amplitude for a single
nanocylinder $D~=~4$ nm illuminated at the plasmonic resonance
($\lambda~=~362$~nm). The same for MM nanotips made of
nanocylinders. (b) $\Lambda~=~1.1~D$, $\lambda~=~331~$nm;
(c) $\Lambda~=~1.19~D$, $\lambda~=~574~$nm;
(d) $\Lambda~=~3.5~D$, $\lambda~=~407~$nm.} \label{fig4}
\end{center}
\end{figure}

To sum up, we have suggested a new metamaterial structure, namely a MM nanotip,
that can evoke field enhancement in comparatively large and controllable
spatial domains and for molding the flow of light. Three operational regimes
have been identified that permit for either field concentration inside the tip,
directly at the apex or the formation of a highly subwavelength photonic
nanojet emerging from the tip apex. The very basis of all these effects is the
strong dispersion of the effective permittivity of the MM that forms the tip.
Variation of the period and/or the shape of the nanocylinder array allow to
control the size, shape and location of the hot spot. Based on these
preliminary 2D simulations one can expect  a rather strong field enhancement
for 3D MM nanotips in comparatively large and controllable spatial domains.

\end{document}